\documentclass[12pt]{article}

\def\reft#1{\mbox{Table~\ref{#1}}}

\begin{document}

\centerline{\large\bf On Prediction of EOP}
\bigskip
\centerline{\bf Z. Malkin and E. Skurikhina}
\footnotetext{Communications of the Institute of Applied Astronomy RAS, 1996, No. 93}

\vskip 10mm
\centerline{\bf Abstract}

\medskip
Two methods of prediction of the Pole coordinates and TAI-UTC were tested --
extrapolation of the deterministic components and ARIMA.
It was found
that each of these methods is most effective for certain length of prognosis.
For short-time prediction ARIMA algorithm yields more accurate prognosis,
and for long-time one extrapolation is preferable.
So, the combined algorithm is being used in practice of IAA EOP Service.
The accuracy of prognosis is close to accuracy of IERS algorithms.
For prediction of nutation the program $KSV_-1996_-1$ \ by T. Herring is being used.

% introd.tex

\section{Introduction}

Organizing the IAA EOP Service
we faced the problem of development
practical technique of predicting EOP, necessary to solve
some problems of coordinate and time maintenance.
In the present work results of the first stage of this work are presented.

The methods of predicting EOP are being advanced during
long-duration time by many authors.
General idea of all methods consists in determination of the statistical
characteristics of an EOP series
for period immediately previous to an interval of prediction
and extrapolation of the found statistical characteristics in future.
The methods of prediction may be separated into two groups:

- {\em Deterministic methods}, at application of which an EOP series
   is being modelled by a set of trends, garmonics, polynomials or more
   complex components
   \cite {Ryhlova90,Fong85,McCarthy88,McCarthy91a,Zhu82}.

- {\em Stochastic methods}, which are based on application of
   such models as ARIMA, Least Square Collocation and other
   \cite {Choli91,Feissel88,Hozakowski90,Kosek92,Kosek93,Kosek95,%
   McCarthy88,SPetrov95a,SPetrov95b,Ulrich73}. As a rule, the stochastic
   modeling is being applied not to initial EOP series but to residuals after
   exception of some deterministic components.

Unfortunately, direct comparison of various methods of predicting EOP
and, accordingly, choice of algorithm the most suitable for practice
using only published results is practically impossible
because authors usually use different methods of testing
and different EOP series to assess accuracy. 
Besides the practical algorithm must
meet the requirement of non-interactive use and it should be also validated.

So, we undertaken some work on uniform testing of various algorithms of
prediction. The technique developed with this purpose provides an assessment
of prediction accuracy based on three criteria:

- root mean square residuals predicted of observed values of EOP;

- maximum residual predicted of observed values of EOP;

- influence of possible errors of last observed EOP on the result of prediction
  (it is important for real practice of EOP service
  when accuracy of last operative determinations is usually lower than
  accuracy of more "old" data).

The series EOP(IERS)C04 was used for testing on an interval 1990-1995
and on more short subintervals.

All told above is concerned to prediction of Pole coordinates and
universal time. As for prediction of nutation, the modern models
can provide high accuracy prognosis for long time intervals.

At the first stage of our work extrapolation of deterministic components,
consisting of set of polynomial trend and garmonics, and ARIMA method
were tested. The resulting algorithms, described below, are being routinely
used in IAA beginning from June 1996. As software realizing other methods
of EOP prediction is available their testing will be performed and
routine technique will be corrected if necessary.
% xy.tex

\section{Prediction of the Pole coordinates}

The first method of predicting Pole motion tested in IAA was
the extrapolation of simple model consisting from
trend, Chandler, annual, and semiannual garmonics.
The base interval used for determination of parameters of trend
and garmonics was varied from 1 to 6 years.
The degree of polynomial trend was varied from 1 to 3. It was found
that the best result can be obtained using linear trend with 1000 days
base interval.
To make the first predicted point agree with the last observed
one the method similar to NEOS method \cite{McCarthy91a} is being used
except that weight parameter 180 days was found more adequate than
190 days used in NEOS practice.

Further investigations had showed that there is certain problem at the
bound between observed and predicted series because of the resulting
curve inflects at the first points of predictions. It can disturb
the procedure of interpolation of EOP if method like spline is used.
Besides it was found that ARIMA method provides more accurate result
for short-time prediction. Various combination of parameters of
autoregression and moving average was tested and optimal strategy was
adopted. It should be mentioned that ARIMA procedure is being applied
to the residuals after removing linear trend and three garmonics
mentioned above.
The base interval equal to 1000 days was found to be most adequate for
fitting of deterministic part in this case, too.
For realization of ARIMA method some routines of V. Choli was used.
To get the best accuracy the correction of ARIMA parameters
is being made using maximum likelihood method \cite{Choli91}.

So, the combining algorithm was accepted as final at the moment.
The interval of prediction is being divided into three part:

-- for interval up to 30 days the ARIMA(1,5) method is being used;

-- for interval 30--90 days the ARIMA(1,2) method is being used;

-- for interval greater than 90 days the extrapolation is being used.

\smallskip
Since we use three various algorithm for different length of prognosis
we need some procedure to join three series of predicted coordinates
to one series without jumps and inflections as good as possible.
This procedure consists of two step:

-- the linear trend is being added to each following prognosis to bring
   its first point into agreement with the last point of previous one
   saving the last point of following prognosis as computed;

-- the predicted value for bound point between two prognosis
   is being replaced by average of
   three values: last but one of previous prognosis, bound point and
   the second point of following prognosis, to make the boundary more
   smooth.

The rms differences between predicted and observed EOPs at the interval
1990--1995 (with 10 days step)
and its subintervals are presented in the \reft{tab:pred_xy_acc} (in mas).
The maximum differences (absolute values)
between predicted and observed EOPs are presented
in the \reft{tab:pred_xy_max}. We considered last values as
guaranteed accuracy of prediction.

\begin{table}[ht]
\centering
\caption{Rms differences between predicted and observed Pole coordinates.}
\medskip
\begin{tabular}{|l|rrrrrrrrr|}
\hline
&\multicolumn{9}{|c|}{Days in future} \\
\cline{2-10}
& \multicolumn{1}{c}{10} & \multicolumn{1}{c}{20} & \multicolumn{1}{c}{30} & \multicolumn{1}{c}{40} & \multicolumn{1}{c}{60}
& \multicolumn{1}{c}{90} & \multicolumn{1}{c}{120} & \multicolumn{1}{c}{150} & \multicolumn{1}{c|}{180} \\
\hline
&&&&&&&&&\\ [-4mm]
&\multicolumn{9}{|c|}{1990--1995} \\ [-5mm]
&&&&&&&&&\\
$X_P$ & 3.9 & 7.0 & 9.4 & 11.3 & 14.5 & 17.9 & 18.5 & 18.2 & 18.0 \\
$Y_P$ & 3.0 & 5.8 & 8.6 & 11.2 & 15.8 & 20.8 & 24.7 & 26.7 & 27.6 \\
&&&&&&&&&\\ [-4mm]
&\multicolumn{9}{|c|}{1992--1995} \\ [-5mm]
&&&&&&&&&\\
$X_P$ & 3.8 & 6.7 & 9.0 & 10.7 & 13.6 & 16.7 & 17.2 & 17.4 & 17.7 \\
$Y_P$ & 2.8 & 5.3 & 7.7 &  9.9 & 13.7 & 18.2 & 22.9 & 26.6 & 29.0 \\
&&&&&&&&&\\ [-4mm]
&\multicolumn{9}{|c|}{1994--1995} \\ [-5mm]
&&&&&&&&&\\
$X_P$ & 3.9 & 6.8 & 8.9 & 10.3 & 13.0 & 16.3 & 16.5 & 16.9 & 18.0 \\
$Y_P$ & 2.9 & 5.6 & 8.2 & 10.5 & 14.8 & 19.3 & 25.9 & 31.0 & 34.3 \\
\hline
\end{tabular}
\label{tab:pred_xy_acc}
\vskip 5mm
\caption{Maximum errors of predicted Pole coordinates.}
\medskip
\begin{tabular}{|l|rrrrrrrrr|}
\hline
&\multicolumn{9}{|c|}{Days in future} \\
\cline{2-10}
& \multicolumn{1}{c}{10} & \multicolumn{1}{c}{20} & \multicolumn{1}{c}{30} & \multicolumn{1}{c}{40} & \multicolumn{1}{c}{60}
& \multicolumn{1}{c}{90} & \multicolumn{1}{c}{120} & \multicolumn{1}{c}{150} & \multicolumn{1}{c|}{180} \\
\hline
&&&&&&&&&\\ [-4mm]
&\multicolumn{9}{|c|}{1990--1995} \\ [-5mm]
&&&&&&&&&\\
$X_P$ &13.6 &22.0 &30.6 & 37.7 & 42.0 & 44.7 & 43.9 & 39.9 & 38.5 \\
$Y_P$ &10.0 &17.5 &26.7 & 32.1 & 41.8 & 53.2 & 59.6 & 56.0 & 62.0 \\
&&&&&&&&&\\ [-4mm]
&\multicolumn{9}{|c|}{1992--1995} \\ [-5mm]
&&&&&&&&&\\
$X_P$ &13.6 &19.0 &30.6 & 37.7 & 42.0 & 44.2 & 43.9 & 39.9 & 38.5 \\
$Y_P$ & 8.0 &12.7 &18.8 & 24.1 & 29.9 & 39.9 & 45.6 & 56.0 & 62.0 \\
&&&&&&&&&\\ [-4mm]
&\multicolumn{9}{|c|}{1994--1995} \\ [-5mm]
&&&&&&&&&\\
$X_P$ &14.8 &18.6 &23.5 & 26.2 & 28.8 & 37.0 & 44.1 & 35.8 & 31.5 \\
$Y_P$ & 8.1 &11.6 &16.6 & 21.7 & 29.8 & 36.4 & 46.1 & 56.2 & 62.4 \\
\hline
\end{tabular}
\label{tab:pred_xy_max}
\vskip 5mm
\caption{Influence of errors of the last values on prediction results.}
\medskip
\begin{tabular}{|c|rrrrrrrr|}
\hline
Test&\multicolumn{8}{|c|}{Days in future} \\
\cline{2-9}
& \multicolumn{1}{c}{1} & \multicolumn{1}{c}{3} & \multicolumn{1}{c}{5} & \multicolumn{1}{c}{10} & \multicolumn{1}{c}{20}
& \multicolumn{1}{c}{30} & \multicolumn{1}{c}{60} & \multicolumn{1}{c|}{90} \\
\hline
$1$ & 2.6 & 5.9 & 7.1 & 6.8 & 6.2 & 4.2 & 2.3 & 1.9 \\
\hline
$2$ & 1.2 & 2.2 & 2.2 & 2.1 & 1.9 & 1.7 & 1.4 & 1.2 \\
\hline
\end{tabular}
\label{tab:pred_xy_dist}
\end{table}

The last test had been performed to evaluate how does prediction result
react on errors of last observed values. Two kind of artificial errors
was applied to real observed points:

{\em Test \/} 1: \ the value of 1 mas was added to (or subtract from)
   the EOPC04 value corresponding to the last observed epoch;

{\em Test \/} 2: \ the values of 0.5, 1.0, 1.5 mas were added to (or subtract
   from) the EOPC04 value corresponding to the three last observed epoch;

\smallskip
The test was used only for ARIMA
method because its influence on the extrapolation of trend-garmonics
model can be easily foreseen without special calculations.
Corrections were applied to initial Pole coordinates on the
various places of Pole motion curve. The results were very similar
and typical differences between predictions of real and distorted
EOPC04 series are presented in the \reft{tab:pred_xy_dist}. The
table contains the result of positive correction. For negative
ones the values in the table will be negative, too.

So, one can see that serious degradation of accuracy may occur when
ARIMA method is used for erroneous observed EOP values. It should be mentioned
that this effect practically linearly depends on the value of error.

\section{Prediction of the UT1}

The similar tests to ones performed for predicting Pole coordinates were made
for TAI-UT1. It was found that autoregression method of order 10 is most
accurate for short-time prediction (up to 15 days)
and method used in NEOS \cite{McCarthy91a}
yields the best result for more long-time one.
Before applying ARIMA algorithm the linear trend, annual and semiannual
garmonics, and tidal variations are being removed from initial series.
Before applying NEOS extrapolation algorithm the quadric trend, annual and
semiannual garmonics, and tidal variations are being removed from
initial series.
Base interval used to obtain parameters of trend and garmonics is
equal to 1500 days in both cases.
Combining of two prognosis is being made as described above for
Pole coordinates.

The rms differences between predicted and observed values at the interval
1990--1995 (with 10 days step)
and its subintervals are presented in the \reft{tab:pred_ut_acc}
(in 0.001 sec).
The maximum differences (absolute values)
between predicted and observed values are presented
in the \reft{tab:pred_ut_max}.

\begin{table}[ht]
\centering
\caption{Rms differences between predicted and observed TAI-UT1.}
\medskip
\begin{tabular}{|rrrrrrrrr|}
\hline
\multicolumn{9}{|c|}{Days in future} \\
\hline
\multicolumn{1}{|c}{10} & \multicolumn{1}{c}{20} & \multicolumn{1}{c}{30} & \multicolumn{1}{c}{40} & \multicolumn{1}{c}{60}
& \multicolumn{1}{c}{90} & \multicolumn{1}{c}{120} & \multicolumn{1}{c}{150} & \multicolumn{1}{c|}{180} \\
\hline
&&&&&&&&\\ [-3mm]
\multicolumn{9}{|c|}{1990--1995} \\ [-4mm]
&&&&&&&&\\
 0.9 & 2.6 & 4.3 &  6.0 &  9.2 & 13.5 & 17.7 & 22.7 & 28.7 \\
&&&&&&&&\\ [-3mm]
\multicolumn{9}{|c|}{1992--1995} \\ [-4mm]
&&&&&&&&\\
 0.9 & 2.6 & 4.3 &  5.9 &  8.7 & 12.9 & 17.3 & 23.2 & 30.1 \\
&&&&&&&&\\ [-3mm]
\multicolumn{9}{|c|}{1994--1995} \\ [-4mm]
&&&&&&&&\\
 0.7 & 2.3 & 4.3 &  6.0 &  8.3 & 10.9 & 14.6 & 21.5 & 30.0 \\
\hline
\end{tabular}
\label{tab:pred_ut_acc}
\vskip 5mm
\caption{Maximum errors of predicted TAI-UT1.}
\medskip
\begin{tabular}{|rrrrrrrrr|}
\hline
\multicolumn{9}{|c|}{Days in future} \\
\hline
\multicolumn{1}{|c}{10} & \multicolumn{1}{c}{20} & \multicolumn{1}{c}{30} & \multicolumn{1}{c}{40} & \multicolumn{1}{c}{60}
& \multicolumn{1}{c}{90} & \multicolumn{1}{c}{120} & \multicolumn{1}{c}{150} & \multicolumn{1}{c|}{180} \\
\hline
&&&&&&&&\\ [-3mm]
\multicolumn{9}{|c|}{1990--1995} \\ [-4mm]
&&&&&&&&\\
 2.6 & 9.0 &15.0 & 17.2 & 22.2 & 29.8 & 39.3 & 48.7 & 58.8 \\
&&&&&&&&\\ [-3mm]
\multicolumn{9}{|c|}{1992--1995} \\ [-4mm]
&&&&&&&&\\
 2.6 & 6.2 &11.1 & 14.1 & 20.0 & 29.8 & 39.3 & 48.7 & 58.8 \\
&&&&&&&&\\ [-3mm]
\multicolumn{9}{|c|}{1994--1995} \\ [-4mm]
&&&&&&&&\\
 1.9 & 5.4 &10.8 & 13.8 & 20.0 & 27.4 & 40.3 & 49.1 & 58.2 \\
\hline
\end{tabular}
\label{tab:pred_ut_max}
\end{table}

Authors of \cite{McCarthy91a} recommended to use smoothing of initial
data before extrapolation with "moving degree" of smoothing.
We could not find any significant improvement when using this recommendation.
Possibly, supplement tests are needed to test such possibility of
improvement of prognosis.

Many attempts was made also to improve the prediction algorithm allowing
for empirically found garmonics, but without definite success.

\section{Prediction of nutation}

The appropriate model is apparently the best way to forecast the nutation
angles. Now we use for prediction the model of
T. Herring realized in his program \verb"KSV_1996_1".
To test the program we compared the differences
between this model and IAU80 theory of nutation with EOP(IERS)C04 series. 
The comparison showed that the bias exist
between Herring's model and IERS values
\[ d\psi(Herring) - d\psi(IERS) = 42.61 \ mas \]
\[ d\varepsilon(Herring) - d\varepsilon(IERS) = 4.95 \ mas \] \\
for interval MJD=46000--50100 without substantial slope.
After allowing for the bias the rms differences between \verb"KSV_1996_1"
and EOP(IERS)C04 are equal to 0.59 mas for $d\psi$ and  0.27 mas 
for $d\varepsilon$.
So, Herring's model can be successively used for prediction of nutation.
% conclus.tex

\section{Conclusion}

The two methods of predicting EOP have been tested -- extrapolation of
trend-garmonics model and ARIMA. To get the best result of prognosis
the combining method is being used for routine processing in the IAA EOP
Service. The accuracy of prediction is close to one of IERS algorithms.
So, we can hardly expect the real improvement of prediction with this
class of models. Some authors propose the new stochastic
\cite{Hozakowski90,Kosek95,SPetrov95a,SPetrov95b} and deterministic
\cite{Ryhlova90} methods and promise the significantly better accuracy
than existing algorithms can provide. But to assess their quality
it would be very important to compare them with methods routinely
used now using uniform testing procedure.
\eject

\end{document}